\documentclass[12pt,preprint]{emulateapj}

\begin{document}

\title{The Role of Multiplicity in Disk Evolution and Planet Formation}

\author{
Adam L. Kraus\altaffilmark{1}, 
Michael J. Ireland\altaffilmark{2}, 
Lynne A. Hillenbrand\altaffilmark{3},
and Frantz Martinache\altaffilmark{4}
}

\altaffiltext{1}{Hubble Fellow; Institute for Astronomy, University of
Hawaii, 2680 Woodlawn Dr., Honolulu, HI 96822, USA}
\altaffiltext{2}{Sydney Institute for Astronomy (SIfA), School of Physics,
University of Sydney, NSW 2006, Australia}
\altaffiltext{3}{California Institute of Technology, Department of
Astrophysics, MC 249-17, Pasadena, CA 91125, USA}
\altaffiltext{4}{National Astronomical Observatory of Japan, Subaru
Telescope, Hilo, HI 96720, USA}

\begin{abstract}

The past decade has seen a revolution in our understanding of protoplanetary disk evolution and planet formation in single star systems. However, the majority of solar-type stars form in binary systems, so the impact of binary companions on protoplanetary disks is an important element in our understanding of planet formation. We have compiled a combined multiplicity/disk census of Taurus-Auriga, plus a restricted sample of close binaries in other regions, in order to explore the role of multiplicity in disk evolution. Our results imply that the tidal influence of a close ($\la$40 AU) binary companion significantly hastens the process of protoplanetary disk dispersal, as $\sim$2/3 of all close binaries promptly disperse their disks within $\la$1 Myr after formation. However, prompt disk dispersal only occurs for a small fraction of wide binaries and single stars, with $\sim$80\%--90\% retaining their disks for at least $\sim$2--3 Myr (but rarely for more than $\sim$5 Myr). Our new constraints on the disk clearing timescale have significant implications for giant planet formation; most single stars have 3--5 Myr within which to form giant planets, whereas most close binary systems would have to form giant planets within $\la$1 Myr. If core accretion is the primary mode for giant planet formation, then gas giants in close binaries should be rare. Conversely, since almost all single stars have a similar period of time within which to form gas giants, their relative rarity in RV surveys indicates either that the giant planet formation timescale is very well-matched to the disk dispersal timescale or that features beyond the disk lifetime set the likelihood of giant planet formation.

\end{abstract}


\section{Introduction}

Observational and theoretical advances over the past decade have dramatically improved our understanding of disk evolution and planet formation. Mid-infrared observations with the Spitzer Space Telescope have yielded a comprehensive census of disks in many nearby star-forming regions \citep{Carpenter:2006hf,Hernandez:2007zu,Rebull:2010xf}, and their SEDs and spectra have allowed for detailed studies of their structure \citep{Calvet:2005xf,Espaillat:2007rq,Currie:2009bw,Merin:2010zz} and composition \citep{Furlan:2006nl,Pascucci:2008uq}. Contemporary advances in submillimeter and millimeter observations have revealed the mass and size distributions of disks \citep{Andrews:2005qf,Mann:2010ve}, including the first direct detections of gaps that might be attributed to planet formation \citep{Andrews:2008fu,Hughes:2009tw,Brown:2009sh}. Finally, theoretical modeling has begun to provide new context for disk formation, evolution, and destruction \citep{Alexander:2006sb,Clarke:2009bs,Alexander:2009oo}.

These advances have begun to paint a consistent and detailed picture of disk evolution. Disks initially form early in the Class 0/I protostellar stages \citep{Shu:1987it,Enoch:2009mp}, acting as the conduits through which massive circumstellar envelopes are accreted onto protostars \citep[e.g.,][]{Bate:1997kx}. These disks are thought to have masses similar to that of the central protostar, and hence are likely an active site of disk fragmentation to form binary companions \citep{Bonnell:2001jl,Clarke:2009bs}. After the circumstellar envelopes are depleted, the protostellar disk gradually accretes the rest of its own mass onto the primary star \citep[e.g.,][]{Gullbring:1998pt,Herczeg:2008dv} over an interval of several million years \citep{Haisch:2001om,Hernandez:2007zu,Hillenbrand:2008pd}, evolving to become a ``T Tauri'' or protoplanetary disk. During this time, dust grains in the disk coagulate and settle toward the midplane \citep[e.g.,][]{Weidenschilling:1977lr} and gradually accumulate into larger planetesimals \citep{Lissauer:1993kx,Pollack:1996dk}. Finally, the disk is dispersed after an age of several million years by some combination of accretion \citep{Gullbring:1998pt}, grain growth \citep{Dullemond:2005fk}, planet formation \citep{Pollack:1996dk}, and photoevaporation \citep{Alexander:2006sb}.

Our improved understanding of protoplanetary disk evolution has also provided new context and tests for planet formation models. The two canonical modes proposed for gas giant formation are the assembly of a rock-ice core that can accrete and hold gas \citep[``core accretion'';][]{Pollack:1996dk} and the direct fragmentation and collapse of material in a Toomre-unstable disk \citep[``disk instability'';][]{Toomre:1964fr,Boss:2001sd}. Both models predict the formation of gas giant planets, but with different population features (i.e., semimajor axes and compositions). Most significantly for this work, the two models also predict different planet formation timescales. Disk instability should be most efficient at early times \citep[$\tau \la 0.5$ Myr][]{Boss:2001sd} when the disk retains most of its initial mass. By contrast, core accretion should produce most of its planets at late times since the assembly of a 10-20 $M_{\earth}$ core seems to require several Myr \citep[e.g.,][]{Hubickyj:2005qy,Dodson-Robinson:2008pb}. The characteristic lifetime for circumstellar disks could determine whether core accretion succeeds in producing giant planets, and the epoch of first appearance for exoplanets (or indirect signatures of their existence, like disk gaps) could provide a critical new test for distinguishing between the two suggested modes of planet formation.

The high frequency of multiple star systems \citep[e.g.,][]{Kraus:2008zr,Kraus:2011qy} and the potentially dramatic dynamical effect of binary companions on disks mean that we must determine the relationship between disks and binary companions in order to fully understand disk evolution and planet formation. Binary companions are expected to truncate disks at $\sim$1/2--1/3 and/or $\sim$2--3 times the binary semimajor axis \citep{Artymowicz:1994ir,Beust:2005gc,Kohler:2011lr,Nagel:2010pm,Andrews:2010ec}, so disks in or around binary systems should show significantly different structure than those around single stars. As we showed for the case of CoKu Tau/4 \citep{Ireland:2008kx}, the large inner gap caused by a short-period binary system could masquerade as a signpost for planet formation, indicating that binary vetting is required for the so-called ``transitional disk'' systems.

The rates of grain growth and dust settling are indistinguishable between single stars and wider binary systems  \citep{Pascucci:2008uq}, but the detailed disk properties of closer binaries (with semimajor axes on the order of the characteristic disk size) have not been studied due to the lack of a suitable sample. The dynamically active environment within such binary systems could affect the lifetime of disks by enhancing accretion or dynamically dispersing the disk. In the most extreme case, a binary companion could inhibit disk formation entirely. Protostellar material accretes from the envelope onto the disk at characteristic radii of $\sim$50 AU \citep{Watson:2007oq}, but if the binary dynamically clears the disk at the typical accretion radius and acts as a sink for angular momentum, then material might accrete ballistically directly onto the star \citep{Bate:1997yq} rather than via the disk \citep{Bate:1997kx}.

Previous surveys have found some evidence for a correlation between binarity and disk properties \citep{Jensen:1996uq,Ghez:1997tg,White:2001jf,Cieza:2009fr,Duchene:2010cj}, but the resolution limits of these surveys only included a restricted range of binary separations, with many of the closest binaries falling inside the inner working angles. We recently completed a new, more complete survey of multiplicity in the benchmark star-forming region Taurus-Auriga \citep{Kraus:2011qy}. This survey exploited nonredundant aperture-mask interferometry to achieve unprecedented resolution ($\sim$15-20 mas; 2--3 AU) and sensitivity ($\Delta K \sim$5--6 at 40 mas; $\sim$7--15 $M_{Jup}$ at $\sim$6 AU). 

In this paper, we explore the implications of our new binary census for the well-studied disk population in Taurus-Auriga. In Section 2, we describe our sample and the disk census for solar-type stars in Taurus, and in Section 3, we describe a corresponding sample of ``close binaries'' compiled for other star-forming regions. In Section 4, we combine the disk and binary censuses to explore the impact of binary companions on protoplanetary disks. Finally, in Sections 5 and 6, we discuss the implications of our results for the timescales and outcome of disk evolution and planet formation.

\section{A Circumstellar Disk Census of Taurus-Auriga}

Taurus-Auriga is one of the nearest and best studied star-forming regions, and our knowledge of its circumstellar disk population is generally considered the most comprehensive for any star-forming region. As we showed in \citet{Kraus:2011qy}, its proximity also makes Taurus the best case for a large-scale study of multiplicity in a mainly class II/III population. The combination of these two features offers an unprecedented opportunity to study the role of multiplicity in protoplanetary disk evolution and planet formation. Our preliminary results \citep{Ireland:2008kx} have demonstrated that disk properties could show interesting correlations with binary properties.

The sample we consider in this study is identical to the full sample from \citet{Kraus:2011qy}; 156 stars with spectral types of G0--M4, of which 133 are known to be binary systems with separations of $\la$500 AU or have been surveyed for multiplicity down to projected separations of $\la$7 AU, typically with sufficient contrast to identify any stellar companion outside this limit. For systems with separations of $\ga$500 AU, we consider the binary secondary independently as long as it falls within the spectral type range of interest. We chose this limit because most components of wider systems have high-resolution multiplicity information for both components, whereas in many closer systems, the secondary has not been thoroughly studied (due to technical limitations). As we show in Section 4, there is no evidence that binarity affects disk evolution in this large separation range, but many of these wide pairs could be hierarchical multiples that can contribute to our sample size of close binaries.

The spectral type range of our sample corresponds to masses of $M\sim$0.25--2.5 $M_{\sun}$ \citep{DAntona:1997li,Baraffe:1998yo} and was set by the upper end of the Taurus mass function (which contains only a handful of intermediate-mass stars) and the flux limit of previous high-resolution imaging surveys ($R\sim$15 for adaptive optics observations). We defer to \citet{Kraus:2011qy} for a full discussion of biases and completeness, but here note that given the meager yield of $\le$M4 stars among recent surveys \citep{Scelsi:2008my,Luhman:2009wd,Rebull:2010xf,Findeisen:2010mj}, there is unlikely to be a large population of Taurus members in this mass range remaining to be discovered.

Most optically thick circumstellar disks are easily identified via any number of diagnostic observations. The disk reprocesses stellar flux into thermal emission that spans the infrared and millimeter wavelengths, and mass accretion onto the central star yields UV and optical excesses and high emission line fluxes. However, some disks are much more elusive; a disk with an inner hole will not radiate at shorter infrared wavelengths, and some disks do not accrete at a detectable rate. The (optically thin) emission at millimeter wavelengths is proportional to the dust mass contained in small dust grains, so disks with low masses or larger grains also could fall below the detection limits of most surveys. As such, proving the conclusive absence of circumstellar material is much more difficult than detecting the presence of a disk.

In light of the complications, we have adopted a combination of criteria for deciding whether a disk is present or absent. Any star that has multiple disk signatures (UV excess, broadened H$\alpha$ emission, NIR excess, MIR excess, or millimeter excess) is considered a high-confidence host of a circumstellar disk. However, stars are judged to lack a disk only if mid-infrared observations at 10--30 $\mu$m detect the stellar photosphere and show no evidence of excess emission from cool dust. In practice, the vast majority of diskless stars are identified from Spitzer observations (IRS spectroscopy or MIPS photometry), though a small number of targets have been observed at 10--20 $\mu$m from the ground. 

In Table 1, we summarize the diagnostic observations from past work that contribute to our assessment of disk-bearing status for each Taurus member of spectral type G0--M4. We have assessed these observations in five different wavelength-dependent categories:

\begin{enumerate}

\item Accretion signatures as inferred from a UV excess at $\la$4000 \AA \citep[Balmer jump of $F_{3600\AA}/F_{4000\AA}>0.5$;][]{Herczeg:2008dv}, or broadened H$\alpha$ emission \citep[$v_{10\%}>200$ km s$^{-1}$; e.g.,][]{Natta:2004uq}.

\item NIR excess at 2--8 $\mu$m, based on the position in Spitzer/IRAC color-color diagrams \citep[e.g.,][]{Hartmann:2005fd}.

\item MIR excess at 10--30 $\mu$m, based on the presence of a significant excess in the $K_s-N$ or $K_s-[24]$ colors \citep[e.g.,][]{Rebull:2010xf} or the observed Spitzer-IRS spectrum \citep{Furlan:2006nl} when compared to the expected values for purely photospheric colors or spectra ($S_{\nu}\sim$1--5 mJy).

\item FIR excess at 30-100 $\mu$m, where given current instrument sensitivities, any detection represents a significant excess over photosphere.

\item Millimeter excess at $>$1 mm, where any detection also represents a significant excess over photosphere.

\end{enumerate}

\noindent These criteria correspond to the presence of circumstellar disks at characteristic radii of $\la$0.1 AU (accretion and NIR excesses), $\la$10--50 AU (MIR excess) and $\ga$50 AU (FIR excess). The millimeter excess is proportional to the total dust mass, which is nominally independent of its radial distribution, but affected by evolution of grain properties. Based on the sum of all of these previous assessments, we then make a judgement as to whether each member hosts a disk (or for binary systems, at least one disk, since most binaries do not have high-resolution MIR data). We also report the separation of binary systems, denoting objects with no binary detection down to $\la$50 mas ($\la$7 AU) as single and making no judgement for stars without sufficient high resolution imaging (``...'').

Our strategy could be prone to a systematic uncertainty for circumbinary disks with extremely large inner gaps, as might be expected for 20--40 AU binary systems if they don't also host inner circumstellar disks. Observations show that inner disk radii as wide as $\sim$50 AU allow for sufficient amounts of warm dust for the system to have a 24$\mu$m excess \citep[e.g.,][]{Espaillat:2007rq}, but it is unclear whether extremely wide circumbinary rings (with radii of $\ga$100 AU) would yield 24$\mu$m detections. An even more conservative approach would require photosphere detections at longer wavelengths, but such measurements have not been feasible. We report all available upper limits, and the absence of a significant number of 70$\mu$m, 160$\mu$m, or submm/mm excess sources without 24$\mu$m excesses suggests that this possible systematic uncertainty might not be significant. The only exception is LkHa 332 G1, which has a marginal excess at 850$\mu$m, but upper limits at 450$\mu$m and 1.3mm \citep{Andrews:2005qf}. We suggest that its nature should be investigated in more detail to determine if the marginal excess was real, and that the overall census likely should be revisited once more sensitive surveys (such as with Herschel) become available.

Of the total sample of 156 targets, 23 lack high-resolution imaging information to determine binarity and 7 lack sufficient disk diagnostic information. Another 14 targets have been omitted from the sample because they are Class 0/I systems with significant envelopes that would show the same MIR signatures as circumstellar disks, whether or not disks are actually present; these targets are also difficult to survey for multiplicity, since resolved nebulosity can mask or mimic the presence of binary companions. Our subsequent analysis uses the 120 stars with sufficient information to determine both their multiplicity and their disk-bearing status, so our census should not show any large biases except those tied to the master census of Taurus membership. Out of these 120 targets, we find a raw disk fraction of 83/120 ($69.2^{+3.8}_{-4.5}\%$); 116 of the targets under consideration have observations available in the 10--30 $\mu$m range that we consider critical for a high-confidence assessment of disk-bearing status.

\section{Disk Census for Close Binaries in Other Associations}

The frequency and properties of disks are expected to evolve on a timescale of $\sim$3--5 Myr \citep[e.g.,][]{Haisch:2001om,Hernandez:2007zu}, so a full characterization of disk evolution and dispersal should consider several regions of different ages. Few star-forming regions have been studied as thoroughly as Taurus-Auriga, so we can not analyze the age-dependent disk frequency with the same level of detail. However, as we describe in Section 4, our results for Taurus \citep[$\tau \sim 2 \pm 1$ Myr;][]{Kraus:2009fk} imply that the disk population is depleted by a constant probability at all separations $\la$40 AU, and therefore all binaries in this separation range might evolve similarly at other ages. If this assumption is valid, then we do not necessarily need a complete binary census for regions at these other ages, as long as we can identify enough systems with any separation inside this limit to comprise a statistically significant sample with respect to any age-dependent trends. To this end, we have compiled a disk census for all known close ($\la$40 AU) binary systems in several other young associations: Ophiuchus (Oph), Chamaeleon-I (Cha-I), Upper Sco (USco), the $\eta$ Cha cluster, the TW Hya association (TWA), and the $\beta$ Pic moving group (BPMG). These populations have mean ages ranging from $\la$1 Myr to $\sim$12 Myr.

In Table 2, we report the disk-hosting status of all known close binary pairs in these populations. As for the Taurus disk census in Section 2, we have assessed the existence of a disk, but do not attempt to study the full SEDs in order to determine disk masses or structures. Unlike for Taurus, we do not compile a full census of all diagnostics, but instead rely exclusively on the 20--30$\mu$m regime, which is probed by observations with Spitzer/IRS or Spitzer/MIPS. Most targets in these populations either have a full suite of data available (accretion, NIR, MIR, and submm/mm) or were neglected by all such surveys, so the full compilation would not allow us to include more systems than the 20--30$\mu$m data alone. The 20--30$\mu$m emission from these systems tracks dust out to separations of at least $\sim$50 AU \citep[as seen from modeling for the inner wall of LkCa 15;][]{Espaillat:2007rq}, so this criterion should include all circumstellar disks and any circumbinary disks out to radii of at least this limit.

In Table 2, we also list the ages that we adopt for each population, as well as the reference source. However, there is some unavoidable ambiguity in age-dating of young stars. As we have demonstrated for Taurus \citep{Kraus:2009fk} and numerous other regions \citep{Hillenbrand:2008pd}, when association members are placed on an HR diagram, they have a range of luminosities corresponding to apparent ages spanning $<$1 Myr to $\sim$5 Myr, even though the median age is $\sim$2 Myr. Given that star formation is still ongoing, then we must conclude that the full age range for typical loose associations could be even larger. Any age that we quote for a stellar population must be regarded as a characteristic value and nothing more. Also, the absolute ages depend on the assumed distances \citep[which can occasionally be wrong by a factor of $\sim$2;][]{Dzib:2011vn} and on the calibration of pre-main sequence evolutionary models \citep[e.g.,][]{Mentuch:2008vn,Lawson:2009rt}. However, calibration errors should affect only the relative age sequence, and most of our target populations have at least one member with a direct parallax determination. We therefore suggest that any observed trends with age are robust to within a rescaling of the population ages.

We found that $\eta$ Cha, TWA, and BPMG are too small to individually contribute to our study, with only 3 close binaries in each population. They are also the oldest ($\sim$8--12 Myr), and their ages could plausibly fall anywhere within this range. We therefore combine all three populations into a single group of age 9$\pm$1 Myr for our subsequent analysis. None of the three has more than one disk-bearing binary system, so this choice does not produce any large systematic effects that will bias our results. However, we also note that these associations might be more dynamically evolved \citep[such as $\eta$ Cha; ][]{Murphy:2010lr}, surrounded by an extended halo of stars that have already been removed. It is possible that the systems left in the core regions are systematically less dynamically evolved (and hence, less likely to have had their disks dynamically stripped in close interactions).

Finally, we have compiled a set of disk frequencies (irrespective of binarity, which typically is not well known) for most of the nearby young regions that have been observed with Spitzer, concentrating specifically on the mass range corresponding to our Taurus sample ($\sim$0.25-2.0 $M_{\sun}$). In Table 3, we list the age, mass range (in mass or spectral type), sample size, and disk frequency for 13 young stellar populations. We can use this sample to place our results on the multiplicity-disk correlation in the overall context of all nearby star-forming regions by comparing them to the total (binary and single) disk frequencies for other regions, even for those regions where multiplicity surveys have not yet identified their binary populations.

\section{The Influence of Multiplicity on Protoplanetary Disks}

Our combined census of circumstellar disks and multiple systems offers an unprecedented opportunity to study the impact of multiplicity on the structure, evolution, and population statistics of protoplanetary disks. For example, one of the most exciting developments over the past decade was the rapid increase in the number of circumstellar disks known to have cleared gaps or inner holes. These cleared regions could be signposts of ongoing planet formation, though other processes (i.e., photoevaporation or binary tidal truncation) could also clear portions of the disk. A few such systems have been known since the 1980s \citep{Strom:1989rr,Skrutskie:1990tt}, but there are now several dozen such stars \citep{Forrest:2004mh,Calvet:2005xf,Espaillat:2007rq,Cieza:2010ys}.  These gaps have traditionally been identified from SED modeling and the characteristic absence of warm material to emit at wavelengths of $\la$10 $\mu$m. However, an increasing number of these gaps are being resolved directly with submm/mm imaging.

Several of our Taurus multiplicity survey targets have been suggested to have cleared gaps or inner holes, including GM Aur \citep{Strom:1989rr,Hughes:2009tw}, UX Tau and LkCa 15 \citep{Espaillat:2007rq}, and CoKu Tau/4 \citep{Forrest:2004mh}. As we reported in \citet{Ireland:2008kx}, CoKu Tau/4 is a close binary pair that appears to be clearing its inner disk through tidal truncation. However, none of the other systems appear to have stellar or brown dwarf companions with separations of $\sim$3--50 AU \citep{Kraus:2011qy}, and surveys with long-baseline interferometry have ruled out stellar companions to even smaller separations \citep[0.35-4 AU;][]{Pott:2010mw}. The inner disk radii for these targets are 20--50 AU and binary companions should truncate disks at radii of $\sim$2--3 times the binary semimajor axis \citep{Artymowicz:1994ir}, so our results argue against binary truncation as the primary formation mechanism, though we can not conclusively rule out binarity in each individual case since a companion very near conjunction could have a projected separation smaller than these surveys' inner working angle. Any binary system with a semimajor axis of $\sim$10--20 AU should spend only a small fraction of its orbit with a projected separation of $\la$1 AU (such that it was not detectable in our observations), so a future observation should establish definitive proof that these disk gaps are not a result of binarity.

On the surface, this paucity of cleared circumbinary disks would seem to contradict our results from \citep{Kraus:2011qy}. The binary frequency for separations of $\sim$4--40 AU is $\sim$20\%, and an extrapolation of the separation distribution seen for field binaries \citep{Duquennoy:1991zh,Raghavan:2010sz} implies that there should be many more systems with separations of $\la$3 AU. We therefore should naively expect dozens of Taurus members to host circumbinary disks that mimic the observational signatures of transitional disks. Given that there seem to be only $\sim$5--10 transitional disks in Taurus, circumbinary disks should overwhelm the population of true transitional disks.

The explanation for this contradiction can be seen in Figure 1, where we plot the disk frequency as a function of binary separation for the binary systems in Taurus. The disk frequency for close ($\la$40 AU) binary systems is only $37^{+9}_{-7} \%$, much lower than the disk frequencies for wider binaries ($90^{+3}_{-8} \%$) and for apparently single stars ($80^{+4}_{-6}\%$). There are indeed many close binary systems in Taurus, but the majority seem to have dispersed their circumstellar disks within an interval shorter than the typical age of Taurus members \citep[$\sim$2 Myr;][]{Kraus:2009fk}. Given the harsh dynamical environment around a binary system, this trend is not unexpected, and signs of this trend have been suggested for some time \citep{Jensen:1996uq,Ghez:1997tg,White:2001jf,Cieza:2009fr,Duchene:2010cj}. However, most of these past surveys had inner working angles of only $\sim$100 mas ($\sim$15 AU), so they could only sample the outermost bin of ``close'' binary systems seen in Figure 1. Most of our newly-discovered close binary systems were beyond previous detection limits \citep{Kraus:2011qy}.

 \begin{figure*}
 \epsscale{1.0}
 \plotone{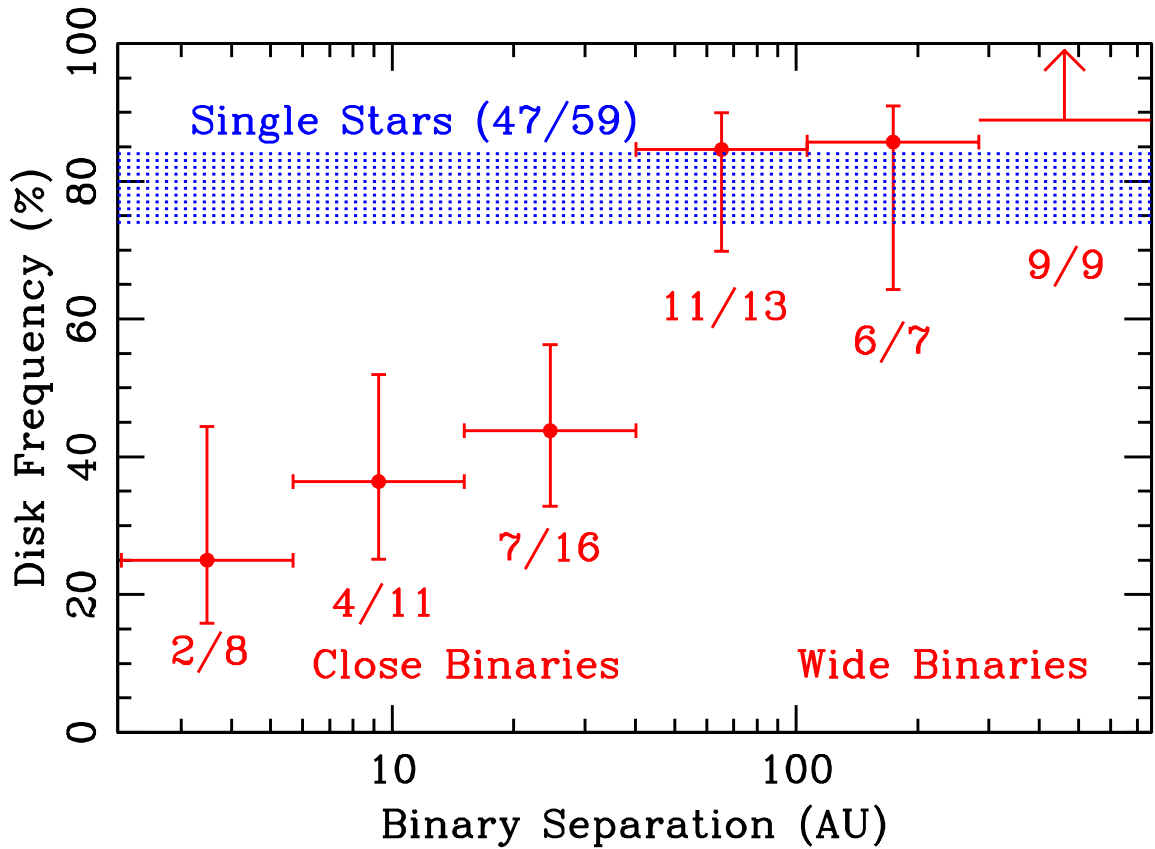}
 \caption{Disk frequency as a function of binary projected separation for G0-M4 stars in the 2 Myr old Taurus-Auriga association. Six ranges of binary separations are shown with red points (where the vertical error bars represent the 1$\sigma$ confidence interval containing the central 68\% of the binomial PDF), while corresponding 1$\sigma$ confidence interval for apparently single stars ($80^{+4}_{-6}\%$) is shown with a blue shaded band. The disk frequency at separations of $\ga$40 AU is indistinguishable from the single-star disk frequency, whereas the disk frequency for close binaries is significantly lower.}
 \end{figure*}

 \begin{figure*}
 \epsscale{1.0}
 \plotone{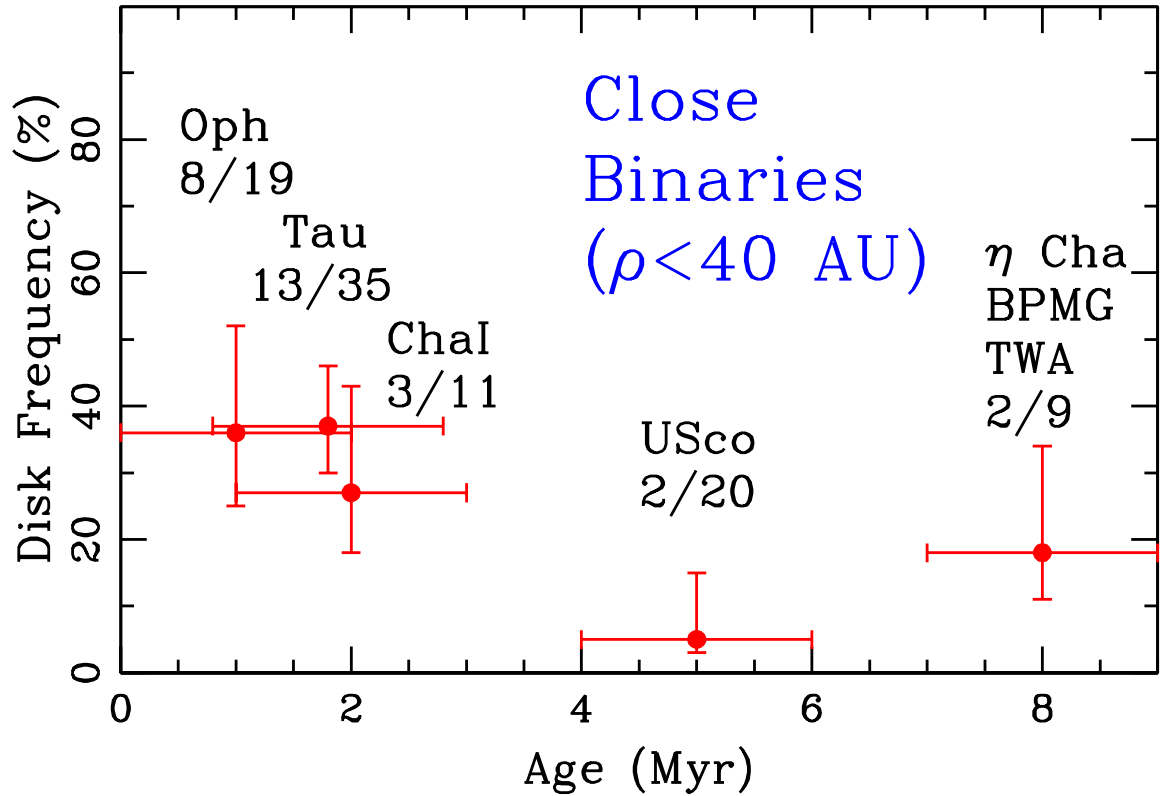}
 \caption{Disk frequency as a function of age for close ($\rho \la 40$ AU) binary systems among G, K, and early M primaries in several nearby young associations. Our results indicate that the majority ($\sim$2/3) of all binary systems lose their protoplanetary disks at ages of $\la$1 Myr, or perhaps never reformed a stable disk after the fragmentation of the companion. However, some binary systems do seem to retain their disk even to ages of $\sim$10 Myr, similar to single stars, as 2/9 binaries in nearby moving groups still maintain optically thick circumbinary disks.}
 \end{figure*}

 \begin{figure*}
 \epsscale{0.8}
 \plotone{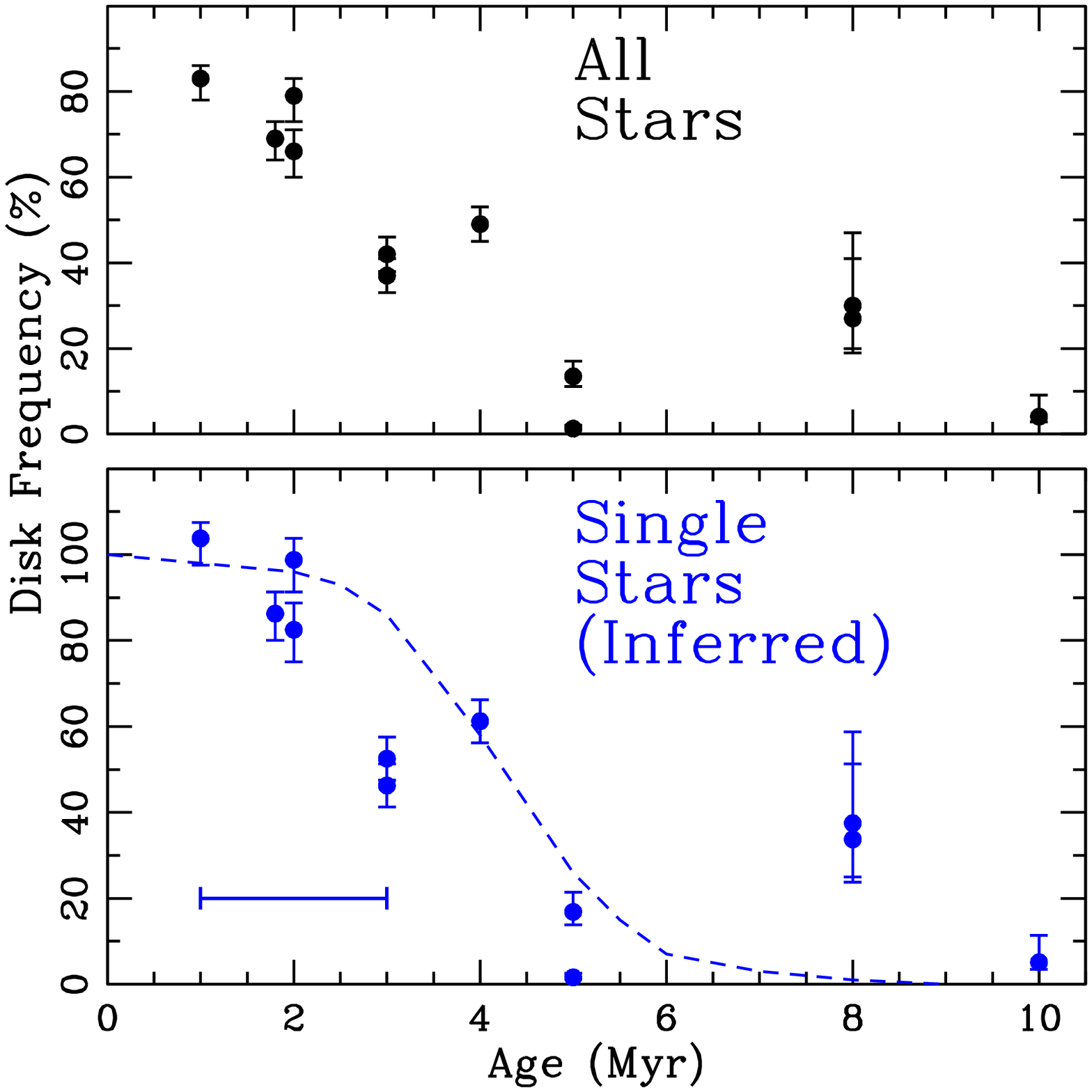}
 \caption{Disk frequency as a function of age for solar-type (G, K, and early M) members of several nearby young associations, showing both the measured frequencies for all members (binary and single; top) and the inferred frequencies for only single stars (bottom). We infer this single-star disk frequency by noting that $\sim$30\% of all solar-type stars have 1--40 AU binary companions \citep{Kraus:2011qy} and that $\sim$2/3 of all close binaries appear to lose their disks very quickly, while the other $\sim$1/3 evolve on a timescale similar to that of the overall population (e.g. Figure 2). Our conclusion is that very few single stars disperse their disks before an age of $\sim$2 Myr, but the vast majority have done so by the age of $\sim$5 Myr. This picture is very similar to the theoretical predictions suggested by \cite{Alexander:2009oo} based on disk evolution models that include both planet formation and photoevaporation; we show their predicted disk frequency as a function of time with a dashed line. The scatter about this relation at young ages is consistent with the uncertainty in ages (at least $\sim$1 Myr, shown in the bottom panel), but the higher disk frequency in the older moving groups appears to be a significant outlier.}
 \end{figure*}

\section{The Timescale for Disk Clearing in Close Binary Systems}

Our results suggest that the majority of close binary systems lose their disks more quickly than single stars, so the next step should be to estimate the clearing timescale by observing many populations of different ages. Thus far, we have surveyed only two populations (Taurus and Upper Sco) with high completeness to separations of $\sim$3 AU, so we can not duplicate this analysis on a large scale. However, the disk frequency is consistent with a uniform value for all Taurus binaries with projected separations of $\la$40 AU, which indicates that we might not need a complete census of each region. If we bin all binaries with projected separations of $\la$40 AU, we can then study the disk frequency of those close binaries in an unbiased manner. 

In Figure 2, we plot the disk frequency as a function of age for the populations described in Section 3 and Table 2. It appears that even at an age of $\sim$1 Myr (for Ophiuchus), the disk frequency for $<$40 AU binary systems has already declined to $\sim$35\%. The disk fraction is similar at the age of $\sim$2 Myr (Taurus and Cha-I), but declines to $5^{+10}_{-2}\%$ at the age of Upper Sco ($\sim$5 Myr). By contrast, the overall disk frequency for the combined single and binary population remains high ($\sim$80--90\%) through the age of Taurus before plummeting by the age of Upper Sco. The inference is that the single star frequency remains even higher than the overall frequency in populations like Ophiuchus and Taurus. The relation between age and disk frequency for binary systems is sampled too sparsely to choose or fit a specific functional form, but the overall shape is similar for both binary systems and single stars, with only a multiplicative factor of $\sim$3 separating them. We therefore suggest that $\sim$2/3 of all close binary systems clear their disks extremely quickly, within $\la$1 Myr of the end of envelope accretion. The other $\sim$1/3 of close binary systems evolve on a timescale similar to that of single stars.

From a theoretical standpoint, it is unclear how quickly binary systems should be expected to lose their circumstellar disks. Observations of Class I protostars suggest that material accretes from the envelope onto the protostellar disk at a characteristic radius of $\sim$50 AU away from the primary star \citep[e.g.,][]{Watson:2007oq}, a distance which likely depends on the characteristic angular momentum left in the envelope. From this point, material then should accrete to the central star via viscous evolution of the disk. However, this picture could be complicated by the presence of a binary companion, as simulations of circumbinary disks suggest that tidal interactions will typically truncate the disk at $\sim$2--3 times the binary semimajor axis \citep{Artymowicz:1994ir}. If the dynamically cleared region extends out to the characteristic radius at which envelope accretion occurs, then there will be no disk left for accreted material to encounter. The process of envelope accretion has not been modeled in the case of a protostellar binary, but it seems plausible that some of the material would accrete ballistically directly down to the central stars \citep{Bate:1997yq}, while the rest would be tidally driven back out toward the envelope. If little or none of the material accretes onto the outer circumbinary disk, then the disk would be cut off from replenishment as it viscously evolves to accrete its own mass down to the central stars.

Even if material in the envelope accretes onto the outer circumbinary disk, the disk could still face a shorter lifetime due to stronger disk dispersal processes. For example, typical circumstellar disks seem to be affected by photoevaporation only at late stages of their evolution, after most of their mass has been accreted to the central star \cite[e.g.,][]{Alexander:2009oo,Alexander:2006sb}. Before this point, the bulk of the disk is self-shielded by the inner edge of the disk at radii of $\la$0.1 AU, which is too deep inside the gravitational radius ($R_G\sim$5--10 AU) to be photoevaporated. However, the inner edge of a circumbinary disk is much higher in the potential well, and so even though the flux of UV photons in the direct radiation field is much lower, those photons would be sufficient to drive disk material completely out of the system. Tides from the central binary could also promote enhanced coalescence of material in the disk, potentially driving enhanced accretion episodes that would drain the disk more quickly down to the central stars via processes like FU Ori outbursts \citep[e.g.,][]{Reipurth:2004pb}. One potential test for this hypothesis would be to study very short-period systems, with periods of days to weeks, since these binaries would interact with most of the disk in a manner similar to single stars. However, the sample of spectroscopic binaries is small and might be biased. Three of the five known SBs in Taurus with $P<1$ yr have disks \citep[][]{Reipurth:1990lr,Mathieu:1997mr,Simon:2000ty,Duchene:2003kn,White:2005gz}, yielding a frequency of $60 ^{+16}_{-22}\%$. The recent discovery of the transiting circumbinary planet Kepler-16 b clearly indicates that gas giant planet formation can occur around short-period systems \citep[][]{2011arXiv1109.3432D}.

It is unclear why a significant fraction of all binary systems ($\sim$1/3) would not be affected by these processes. Indeed, some of the oldest disks in our sample are associated with binary systems (such as HD 98800 and $\eta$ Cha 9), so some disks can persist in close binary systems for as long as 7--10 Myr. The current statistically robust sample of disks is not sufficient for studying the dependence of disk lifetime on additional parameters, but it is plausible that the disk lifetime could depend on the parameters of the binary system, such as the eccentricity and mass ratio. The tidal effects that open different resonances in the disk are sensitive to both eccentricity and mass ratio, with higher-order resonances being opened with high eccentricity or low mass ratio \citep{Artymowicz:1994ir}. We therefore suggest that the most dynamically stable systems (circular, equal-mass binaries) might be the best candidates for long-term disk survival, a hypothesis that should be tested more robustly with larger samples and with additional detailed observations of individual circumbinary disks \citep[e.g.,][]{Jensen:2007zo,Boden:2009lr}

\section{The Frequency and Timescale of Planet Formation}

Our results also have significant implications for the evolutionary history of disks around single stars. Most surveys of the disk frequency as a function of age have suggested that the disk frequency declines as a linear or exponential function over time \citep{Haisch:2001om,Furlan:2006nl,Hernandez:2007zu,Hillenbrand:2008pd,Mamajek:2009lr}. However, none of those surveys were stringently vetted to remove close binary systems, and many of the populations in those studies are too distant to identify binaries with separations of $\la$40 AU. Since these close binaries appear to lose their disks more quickly (Section 5), they will bias the overall disk frequency downward. Alternately, for a given disk frequency in a total sample where $\sim$20--30\% of sample members are close binaries and $\sim$2/3 of those close binaries lose their disks promptly after formation \citep[Section 5;][]{Kraus:2011qy} the corresponding single-star disk frequency should be $\sim$15--20\% higher.

In Taurus, the observed disk frequency for all stars in our study's mass range ($\sim$0.25--2.5 $M_{\sun}$) is $\sim$70\% (Section 2); the corresponding single-star disk frequency after applying this correction should be $\sim$80\%--85\%, as is confirmed by our updated census (Section 4; Figure 1). In contrast, the disk frequencies for Upper Scorpius and NGC 2362 ($\tau$$\sim$5 Myr) are only $\sim$5--10\% in this mass range \citep{Carpenter:2006hf,Dahm:2007nx,Currie:2009ix}. The steep decline across the 2--5 Myr age range suggests that almost all single stars host a circumstellar disk for an interval of $2 < \tau < 5$ Myr, with little need for an immediate dispersal mechanism as suggested by the linear or exponential decline of previous disk surveys \citep[e.g.,][]{Haisch:2001om,Hernandez:2007zu}. A linear or exponential functional form for the single-star disk frequencies in Taurus and Upper Sco would be inappropriate, as the fit would exceed a 100\% disk fraction when extrapolated to zero age.

Our new picture for disk evolution also closely resembles the disk frequency evolution predicted by \citet{Alexander:2009oo}, who simulated the simultaneous effects of giant planet formation and disk photoevaporation to show that the single-star disk frequency should decline abruptly over ages of 2--5 Myr. We illustrate this resemblance in Figure 3, where we show the observed total disk frequency for 13 nearby young populations that have been observed with Spitzer, and then we infer a single-star disk frequency by assuming that $\sim$30\% of all stars have 1--40 AU binary companions \citep{Kraus:2011qy} and that $\sim$2/3 of those close binaries lost their disks very quickly, while the other $\sim$1/3 evolve as single stars do. As can be seen from the inferred single-star disk frequency, this assumption yields excellent agreement with the results of \citet{Alexander:2009oo}, albeit with some long-lived disks in the lowest-density moving groups that remain unexplained by theory.

These conclusions have significant implications for the timescale of giant planet formation in binary systems. If many close ($\la$40 AU) binary systems lose their disk within $\la$1 Myr, then they are unlikely to form giant planets via core accretion; the most recent models suggest that core accretion requires several million years to produce cores massive enough to accrete gas out of the circumstellar disk even near the snow line \citep[$\sim$3--5 AU; e.g.,][]{Hubickyj:2005qy,Dodson-Robinson:2008pb}. This prediction is consistent with tentative results for exoplanet host stars in the field that have been identified via either radial velocities \citep{Desidera:2007yq,Eggenberger:2007vn} or planet transits \citep{Daemgen:2009my}, who have found that some exoplanet hosts have binary companions, but they tend to fall only at separations of $\ga$100 AU. The RV discoveries could be biased by rejection of known binary systems from the target samples, as the spectra of stars in close binaries tend to be affected by spectral contamination from the other component. However, transit searches are not biased to the same degree; similar-brightness binaries might be rejected during followup as indicating probable blends, but faint secondaries should not trigger any rejection, and so they would remain to be identified in subsequent high-resolution imaging \citep{Daemgen:2009my}. This apparent anticorrelation aside, it does appear that some giant planets form in close binary systems \citep{Hatzes:2003rt,Chauvin:2011ys}, so the existence and strength of the effect should be investigated further. Furthermore, planet formation is likely to proceed quite differently in the outer parts of circumbinary disks (at 50--100 AU) than in the inner parts of circum-primary or circum-secondary disks, much less in continuous and viscously evolving disks around single stars. The outcome of planet formation likely depends on the binary properties and resulting disk properties, rather than simply the existence of a binary companion. Finally, we note that if many giant planets are found in future RV surveys that are not as biased against binary systems, then it could provide evidence that a significant fraction of planets form via the quicker process of gravitational instability \citep{Boss:2001sd}.

Conversely, the majority of single stars ($\ga$80\%) appear to host a circumstellar disk until the age of 2--5 Myr, with little evidence for a prompt dispersal mechanism that shuts of planet formation on timescales of $\la$1 Myr. Given that most single stars have a relatively uniform range of disk lifetimes, then they might be expected to have similar opportunities for giant planet formation. We therefore infer that the low frequency of extrasolar gas giant planets \citep[$\sim$10\% for masses $\ga$0.3 $M_{Jup}$ and orbital radii $\la$3 AU;][]{Cumming:2008zr} either is not set by disk dissipation shutting off giant planet formation, or is evidence that the disk dissipation timescale is very well-matched to the giant planet formation timescale \citep[$\sim$3 Myr for core accretion;][]{Pollack:1996dk,Hubickyj:2005qy,Dodson-Robinson:2008pb}. In the latter case, disk dissipation in the latter stages of core growth should leave a large population of super-Earths and mini-Neptunes that were not massive enough to accrete gas ($M\la$20 $M_{Earth}$) waiting to be discovered near the snow line ($\sim$3--5 AU). This hypothesis is supported by the high planet frequency recently reported from microlensing results by \citet{Sumi:2010mz}, who suggest that in the separation range most sensitively probed by microlensing ($\sim$2--6 AU), the mass function is sufficiently steep ($dN/d\log(q) \propto q^{-0.7\pm0.2}$ that Neptune-mass planets that might be failed cores ($M\sim$5--20 $M_{Earth}$) are 6.8$^{+6.6}_{-3.4}$ times as common as gas giant planets ($M\sim$0.3--3.0 $M_{Jup}$). Transit results from Kepler \citep[e.g.,][]{Howard:2011uq} also indicate that small planets (as measured by their radius) are quite common, although the Kepler census does not yet reach the snow line. Finally, the uniformly long lifetime for disks also argues that fast planet formation via gravitational instability \citep{Boss:2001sd} is probably intrinsically rare, or else gas giants would be ubiquitous around single solar-type stars.

\section{Summary}

We have combined a census of all previous disk surveys in Taurus-Auriga and several other nearby star-forming regions with our own previous multiplicity results. Combined with age estimates from the literature, this data set reveals several significant conclusions for protoplanetary disk evolution and planet formation:

\begin{enumerate}

\item The tidal influence of a close ($\la$40 AU) binary companion inhibits the formation or speeds the dispersal of protoplanetary disks. At ages of 1--2 Myr, $\sim$2/3 of all close binaries in Taurus-Auriga have no disk; this trend seems to hold even in younger populations like Ophiuchus, which suggests that most disks in close binary systems might not survive past the exhaustion of their mass reservoirs in the extended circumstellar envelope. However, $\sim$1/3 of all close binaries have disk lifetimes similar to those of single stars, and some close binary systems retain disks for 5--10 Myr. These caveats show that stable disk configurations in binary systems do exist.

\item Once close binary systems are removed from the disk census, we find that $\sim$80\%--90\% of both wide binaries and single stars retain their disk for at least $\sim$1--2 Myr. Only a small fraction of single stars ($\sim$10--20\%) promptly disperse their protoplanetary disks. Our results also show that sample vetting is critical for disk studies, as the frequency for genuinely single stars could be biased downward by as much as $\sim$15--20\%. For regions which are more distant and can't be surveyed for close binaries, this correction must be applied on a statistical basis.

\item Our new constraints on the disk clearing timescale have significant implications for giant planet formation. Most single stars and wide binaries ($80^{+4}_{-6}\%$ and $90^{+3}_{-8}\%$, respectively, in Taurus) retain their disks for $\ga$2 Myr, but few ($\sim$15\%, in Upper Sco) retain them for $\sim$5 Myr. The conclusion is that most single stars and wide binaries have a similar length of time within which to form planets, and thus that the low frequency of giant planet formation ($\sim$5--10\%) can be limited by the lifetime of disks only if the planet formation and disk dispersal timescales are very well-matched. Conversely, a majority of close binaries clear their disks within $\la$1 Myr, and thus it seems unlikely that giant planets could form via core accretion in those systems. Both of these predictions are supported by preliminary results from RV and microlensing planet searches, which find that planets in close binary systems are likely rare and single stars host many Neptune-mass planets (i.e. failed cores) for every gas giant.

\end{enumerate}

\acknowledgements

The authors thank G. Herczeg, S. Andrews, S. Corder, T. Currie, I. Pascucci, and R. Alexander for helpful discussions on envelope and disk evolution, and we thank the referee for providing a thoughtful and helpful critique of our paper. ALK was supported by a SIM Science Study and by NASA through Hubble Fellowship grant 51257.01 awarded by the Space Telescope Science Institute, which is operated by the Association of Universities for Research in Astronomy, Inc., for NASA, under contract NAS 5-26555. MJI was supported by an Australian Postdoctoral Fellowship from the ARC. LAH acknowledges support from the NASA Origins of Solar Systems and the NASA ADP programs.

\bibliographystyle{/users/akraus/Dropbox/Papers/apj.bst}
\bibliography{/users/akraus/Dropbox/Papers/krausbib}

\clearpage

\LongTables

 \begin{deluxetable}{lcccllllll}
 \tabletypesize{\tiny}
 \tablewidth{0pt}
 \tablecaption{Circumstellar Disk Census of Taurus-Auriga}
 \tablehead{\colhead{Name} & \colhead{RA} & \colhead{DEC} & \colhead{Binary} & \colhead{Accretion} & \colhead{2--8 $\mu$m} & \colhead{10--30 $\mu$m} & \colhead{30-200 $\mu$m} & \colhead{1--3 mm} & \colhead{Disk}
 \\
 \colhead{} & \multicolumn{2}{c}{(J2000)} & \colhead{Sep (AU)} & \colhead{Present?} & \colhead{Excess} & \colhead{Excess} & \colhead{Excess} & \colhead{Excess} & \colhead{Present?\tablenotemark{a}}
}
\startdata
IRAS 04016+2610&4 04 43.22&+26 18 54.5&...&Y(1)&Y(2)&Y(3)&...(...)&Y(4)&Class I\\
2M04080782&4 08 07.82&+28 07 28.0&6.4&N(5)&...(...)&...(...)&...(...)&...(...)&...\\
LkCa 1&4 13 14.14&+28 19 10.8&single&N(6)&N(7)&N(8)&...(...)&N(4)&N\\
Anon 1&4 13 27.23&+28 16 24.8&2.2&N(6)&N(9)&N(8)&...(...)&N(4)&N\\
IRAS 04108+2910&4 13 56.40&+29 18 15.0&single&Y(10)&Y(9)&Y(8)&Y(9)&N(11)&Y\\
FM Tau&4 14 13.58&+28 12 49.2&single&Y(12)&Y(7)&Y(8)&Y(9)&Y(4)&Y\\
CW Tau&4 14 17.00&+28 10 57.8&single&Y(12)&Y(7)&Y(8)&Y(9)&Y(4)&Y\\
MHO-1&4 14 26.40&+28 05 59.7&...&Y(13)&Y(2)&Y(2)&...(...)&...(...)&Class I\\
MHO 2&4 14 26.40&+28 05 59.7&7.3&Y(1)&Y(2)&...(...)&...(...)&...(...)&Y\\
MHO 3&4 14 30.55&+28 05 14.7&4.5&Y(13)&Y(9)&Y(8)&Y(9)&...(...)&Y\\
FP Tau&4 14 47.31&+26 46 26.4&single&Y(12)&Y(14)&Y(8)&Y(9)&N(11)&Y\\
CX Tau&4 14 47.86&+26 48 11.0&single&Y(15)&Y(14)&Y(16)&Y(9)&Y(4)&Y\\
LkCa 3&4 14 47.97&+27 52 34.7&69&N(10)&N(7)&N(8)&...(...)&N(4)&N\\
FO Tau&4 14 49.29&+28 12 30.6&22&Y(17)&Y(9)&Y(16)&Y(9)&Y(4)&Y\\
LkCa 4&4 16 28.11&+28 07 35.8&single&N(12)&N(7)&N(8)&...(...)&N(4)&N\\
CY Tau&4 17 33.73&+28 20 46.9&single&Y(12)&Y(18)&Y(8)&Y(9)&Y(4)&Y\\
LkCa 5&4 17 38.94&+28 33 00.5&6.9&N(6)&N(9)&N(8)&...(...)&N(4)&N\\
V410 X-ray 1&4 17 49.65&+28 29 36.3&single&Y(10)&Y(9)&Y(9)&Y(9)&...(...)&Y\\
V410 Anon 24&4 18 22.39&+28 24 37.6&...&...(...)&N(18)&N(9)&...(...)&...(...)&N\\
V410 Anon 25&4 18 29.10&+28 26 19.1&...&...(...)&N(18)&N(9)&...(...)&...(...)&N\\
V410 Tau&4 18 31.10&+28 27 16.2&17.8&N(6)&N(18)&N(8)&...(...)&N(4)&N\\
DD Tau&4 18 31.13&+28 16 29.0&80&Y(17)&Y(7)&Y(8)&Y(9)&Y(4)&Y\\
CZ Tau&4 18 31.59&+28 16 58.5&46&Y(15)&Y(18)&Y(8)&Y(9)&N(4)&Y\\
IRAS 04154+2823&4 18 32.03&+28 31 15.4&single&...(...)&Y(9)&Y(3)&Y(9)&Y(4)&Class I\\
V410 X-ray 2&4 18 34.45&+28 30 30.2&...&...(...)&Y(9)&Y(9)&Y(9)&...(...)&Y\\
V410 X-ray 4&4 18 40.23&+28 24 24.5&...&...(...)&N(18)&N(9)&...(...)&...(...)&N\\
LR 1&4 18 41.33&+28 27 25.0&...&...(...)&Y(18)&Y(9)&Y(9)&...(...)&Y\\
V410 X-ray 7&4 18 42.50&+28 18 49.8&4.6&N(13)&Y(9)&Y(9)&...(...)&...(...)&Y\\
V410 Anon 20&4 18 45.06&+28 20 52.8&...&...(...)&N(18)&N(9)&...(...)&...(...)&N\\
Hubble 4&4 18 47.04&+28 20 07.3&4.1&N(12)&N(18)&N(8)&...(...)&N(4)&N\\
CoKu Tau/1&4 18 51.48&+28 20 26.5&33&Y(1)&Y(18)&Y(3)&Y(9)&Y(4)&Class I\\
HBC 376&4 18 51.70&+17 23 16.6&single&N(6)&N(2)&N(2)&...(...)&N(4)&N\\
FQ Tau&4 19 12.81&+28 29 33.1&109&Y(17)&Y(7)&Y(16)&Y(9)&Y(4)&Y\\
BP Tau&4 19 15.84&+29 06 26.9&single&Y(12)&Y(7)&Y(8)&Y(9)&Y(4)&Y\\
V819 Tau&4 19 26.26&+28 26 14.3&single&N(12)&N(18)&Y(3)&...(...)&N(4)&Y\\
LkCa 7&4 19 41.27&+27 49 48.5&148&N(17)&N(18)&N(8)&...(...)&N(4)&N\\
IRAS 04166+2706&4 19 41.48&+27 16 07.0&...&Y(5)&Y(18)&Y(3)&...(...)&Y(4)&Class I\\
IRAS 04187+1927&4 21 43.24&+19 34 13.3&single&...(...)&Y(2)&Y(8)&...(...)&...(...)&Y\\
DE Tau&4 21 55.64&+27 55 06.1&single&Y(12)&Y(7)&Y(8)&Y(9)&Y(4)&Y\\
RY Tau&4 21 57.40&+28 26 35.5&single&Y(12)&Y(9)&Y(8)&Y(9)&Y(4)&Y\\
HD 283572&4 21 58.84&+28 18 06.6&single&N(19)&N(18)&N(8)&...(...)&N(4)&N\\
Haro6-5B&4 22 00.69&+26 57 33.3&...&Y(1)&Y(9)&Y(9)&Y(9)&...(...)&Class I\\
FS Tau&4 22 02.18&+26 57 30.5&33&Y(17)&Y(9)&Y(8)&...(...)&Y(4)&Y\\
LkCa 21&4 22 03.14&+28 25 39.0&6.4&Y(19)&N(18)&N(8)&...(...)&N(4)&N\\
CFHT-Tau-21&4 22 16.76&+26 54 57.1&single&Y(5)&Y(9)&Y(9)&Y(9)&...(...)&Y\\
FT Tau&4 23 39.19&+24 56 14.1&&...(...)&...(...)&...(...)&...(...)&...(...)&...\\
IRAS 04216+2603&4 24 44.58&+26 10 14.1&single&Y(10)&Y(14)&Y(8)&Y(9)&N(11)&Y\\
IP Tau&4 24 57.08&+27 11 56.5&single&Y(12)&Y(18)&Y(8)&Y(9)&Y(4)&Y\\
J1-4872 A&4 25 17.68&+26 17 50.4&25&N(20)&N(14)&N(9)&...(...)&N(4)&N\\
FV Tau&4 26 53.53&+26 06 54.4&102&Y(17)&Y(7)&Y(8)&Y(9)&Y(4)&Y\\
FV Tau /c&4 26 54.41&+26 06 51.0&102&Y(17)&Y(7)&Y(9)&...(...)&N(4)&Y\\
DF Tau&4 27 02.80&+25 42 22.3&10.6&Y(17)&Y(14)&Y(8)&Y(9)&Y(4)&Y\\
DG Tau&4 27 04.70&+26 06 16.3&single&Y(12)&Y(14)&Y(8)&Y(9)&Y(4)&Y\\
HBC 388&4 27 10.56&+17 50 42.6&single&N(6)&...(...)&N(8)&...(...)&N(4)&N\\
IRAS 04260+2642&4 29 04.99&+26 49 07.3&...&Y(1)&Y(14)&Y(9)&Y(9)&Y(4)&Class I\\
J1-507&4 29 20.71&+26 33 40.7&11.5&N(21)&N(14)&N(9)&...(...)&N(4)&N\\
GV Tau&4 29 23.73&+24 33 00.3&187&Y(1)&Y(9)&Y(9)&Y(9)&...(...)&Class I\\
FW Tau&4 29 29.71&+26 16 53.2&21.9&N(6)&N(14)&Y(9)&...(...)&Y(4)&Y\\
IRAS 04264+2433&4 29 30.08&+24 39 55.1&...&Y(1)&Y(18)&Y(3)&Y(9)&...(...)&Class I\\
DH Tau&4 29 41.56&+26 32 58.3&...&Y(15)&Y(14)&Y(8)&Y(9)&Y(4)&Y\\
DI Tau&4 29 42.48&+26 32 49.3&17.4&Y(15)&N(14)&N(8)&...(...)&...(...)&N\\
IQ Tau&4 29 51.56&+26 06 44.9&single&Y(15)&Y(14)&Y(8)&Y(9)&Y(4)&Y\\
2M04295422&4 29 54.22&+17 54 04.1&...&N(5)&...(...)&...(...)&...(...)&...(...)&...\\
UX Tau B&4 30 04.00&+18 13 49.4&19.7&N(20)&N(7)&N(7)&...(...)&...(...)&N\\
UX Tau&4 30 04.00&+18 13 49.4&390&Y(16)&Y(7)&Y(22)&...(...)&Y(4)&Y\\
FX Tau&4 30 29.61&+24 26 45.0&129&Y(15)&Y(7)&Y(8)&Y(9)&Y(4)&Y\\
DK Tau&4 30 44.25&+26 01 24.5&340&Y(12)&Y(14)&Y(8)&Y(9)&Y(4)&Y\\
IRAS 04278+2253&4 30 50.28&+23 00 08.8&...&Y(1)&Y(9)&Y(3)&Y(9)&Y(4)&Class I\\
ZZ Tau&4 30 51.38&+24 42 22.3&6.1&Y(10)&Y(18)&Y(8)&...(...)&N(4)&Y\\
JH 56&4 31 14.44&+27 10 18.0&single&N(10)&N(14)&Y(9)&...(...)&N(4)&Y\\
V927 Tau&4 31 23.82&+24 10 52.9&39&N(23)&N(18)&N(9)&...(...)&N(4)&N\\
LkHa358&4 31 36.13&+18 13 43.3&single&Y(23)&Y(2)&Y(3)&...(...)&Y(11)&Class I\\
HH30 AB&4 31 37.47&+18 12 24.5&...&Y(1)&Y(2)&Y(3)&...(...)&...(...)&Class I\\
HL Tau&4 31 38.44&+18 13 57.7&...&Y(1)&Y(2)&Y(3)&...(...)&...(...)&Class I\\
XZ Tau&4 31 40.07&+18 13 57.2&43&Y(17)&Y(2)&Y(8)&...(...)&...(...)&Y\\
HK Tau&4 31 50.57&+24 24 18.1&340&Y(12)&Y(18)&Y(8)&Y(9)&Y(4)&Y\\
V710 Tau AB&4 31 57.79&+18 21 38.1&470&Y(10)&Y(7)&Y(8)&...(...)&Y(4)&Y\\
V710 Tau C&4 31 59.68&+18 21 30.5&...&Y(24)&Y(24)&Y(2)&...(...)&...(...)&Class I\\
L1551-51&4 32 09.27&+17 57 22.8&single&N(6)&N(2)&N(8)&...(...)&N(4)&N\\
V827 Tau&4 32 14.57&+18 20 14.7&13.5&N(12)&N(7)&N(8)&...(...)&N(4)&N\\
Haro 6-13&4 32 15.41&+24 28 59.7&single&Y(1)&Y(18)&Y(8)&Y(9)&Y(4)&Y\\
V826 Tau&4 32 15.83&+18 01 38.9&single&N(12)&N(2)&N(8)&...(...)&N(4)&N\\
V928 Tau&4 32 18.86&+24 22 27.1&29&N(10)&N(18)&N(8)&...(...)&N(4)&N\\
GG Tau A&4 32 30.35&+17 31 40.6&35&Y(17)&Y(7)&Y(8)&...(...)&Y(4)&Y\\
FY Tau&4 32 30.58&+24 19 57.3&single&Y(15)&Y(18)&Y(9)&...(...)&Y(4)&Y\\
FZ Tau&4 32 31.76&+24 20 03.0&single&Y(6)&Y(18)&Y(8)&Y(9)&Y(4)&Y\\
UZ Tau&4 32 43.04&+25 52 31.1&510&Y(12)&Y(14)&Y(8)&Y(9)&Y(4)&Y\\
L1551-55&4 32 43.73&+18 02 56.3&single&N(6)&N(2)&N(2)&...(...)&N(4)&N\\
JH 112 A&4 32 49.11&+22 53 02.8&220&Y(10)&Y(9)&Y(9)&Y(9)&Y(4)&Y\\
GH Tau&4 33 06.22&+24 09 34.0&44&Y(17)&Y(7)&Y(8)&Y(9)&Y(4)&Y\\
V807 Tau&4 33 06.64&+24 09 55.0&...&Y(17)&Y(18)&Y(8)&...(...)&Y(4)&Y\\
V830 Tau&4 33 10.03&+24 33 43.4&single&N(12)&N(18)&N(8)&...(...)&N(4)&N\\
IRAS 04301+2608&4 33 14.36&+26 14 23.5&...&...(...)&Y(14)&Y(9)&Y(9)&Y(4)&Y\\
IRAS 04303+2240&4 33 19.07&+22 46 34.2&single&Y(1)&Y(9)&Y(8)&Y(9)&...(...)&Y\\
GI Tau&4 33 34.06&+24 21 17.0&single&Y(12)&Y(18)&Y(8)&...(...)&...(...)&Y\\
GK Tau&4 33 34.56&+24 21 05.8&single&Y(12)&Y(18)&Y(8)&Y(9)&Y(4)&Y\\
IS Tau&4 33 36.79&+26 09 49.2&32&Y(17)&Y(14)&Y(8)&Y(9)&Y(4)&Y\\
DL Tau&4 33 39.06&+25 20 38.2&single&Y(12)&Y(14)&Y(8)&Y(9)&Y(4)&Y\\
HN Tau&4 33 39.35&+17 51 52.4&460&Y(12)&Y(2)&Y(8)&...(...)&Y(4)&Y\\
DM Tau&4 33 48.72&+18 10 10.0&single&Y(15)&Y(7)&Y(25)&...(...)&Y(4)&Y\\
CI Tau&4 33 52.00&+22 50 30.2&single&Y(12)&Y(7)&Y(8)&Y(9)&Y(4)&Y\\
IT Tau&4 33 54.70&+26 13 27.5&350&Y(20)&Y(14)&Y(8)&Y(9)&Y(4)&Y\\
J2-2041&4 33 55.47&+18 38 39.1&61&N(26)&N(2)&...(...)&...(...)&...(...)&...\\
JH 108&4 34 10.99&+22 51 44.5&single&N(10)&N(9)&N(9)&...(...)&N(4)&N\\
HBC 407&4 34 18.04&+18 30 06.7&20&N(19)&N(2)&...(...)&...(...)&N(4)&...\\
AA Tau&4 34 55.42&+24 28 53.2&single&Y(12)&Y(18)&Y(8)&Y(9)&Y(4)&Y\\
HO Tau&4 35 20.20&+22 32 14.6&single&Y(15)&Y(9)&Y(8)&Y(9)&Y(4)&Y\\
FF Tau&4 35 20.90&+22 54 24.2&5.3&N(10)&N(9)&N(8)&...(...)&N(4)&N\\
HBC 412&4 35 24.51&+17 51 43.0&102&N(19)&N(2)&N(2)&...(...)&N(4)&N\\
DN Tau&4 35 27.37&+24 14 58.9&single&Y(12)&Y(18)&Y(16)&Y(9)&Y(4)&Y\\
CoKu Tau/3&4 35 40.94&+24 11 08.8&300&...(...)&Y(18)&Y(8)&Y(9)&N(4)&Y\\
HQ Tau&4 35 47.34&+22 50 21.7&...&Y(19)&Y(9)&Y(8)&Y(9)&Y(4)&Y\\
KPNO-Tau-15&4 35 51.10&+22 52 40.1&...&N(27)&N(9)&N(9)&...(...)&...(...)&N\\
HP Tau&4 35 52.78&+22 54 23.1&single&Y(10)&Y(9)&Y(8)&Y(9)&Y(4)&Y\\
HP Tau-G3&4 35 53.50&+22 54 09.0&4.4&N(10)&N(9)&N(2)&...(...)&...(...)&N\\
HP Tau-G2&4 35 54.15&+22 54 13.5&single&N(10)&N(9)&N(8)&...(...)&...(...)&N\\
Haro 6-28&4 35 56.84&+22 54 36.0&94&Y(1)&Y(9)&Y(9)&...(...)&Y(4)&Y\\
LkCa 14&4 36 19.09&+25 42 59.0&single&N(6)&N(14)&N(9)&...(...)&N(4)&N\\
DO Tau&4 38 28.58&+26 10 49.4&single&Y(12)&Y(14)&Y(8)&Y(9)&Y(4)&Y\\
HV Tau&4 38 35.28&+26 10 38.6&5.2&N(10)&N(14)&N(7)&...(...)&...(...)&N\\
VY Tau&4 39 17.41&+22 47 53.4&96&N(10)&Y(2)&Y(8)&...(...)&N(4)&Y\\
LkCa 15&4 39 17.80&+22 21 03.5&single&Y(15)&Y(7)&Y(22)&Y(9)&Y(4)&Y\\
GN Tau&4 39 20.91&+25 45 02.1&49&Y(28)&Y(14)&Y(8)&Y(9)&Y(4)&Y\\
IC2087IR&4 39 55.75&+25 45 02.0&single&Y(1)&Y(9)&Y(3)&Y(9)&Y(4)&Class I\\
IRAS 04370+2559&4 40 08.00&+26 05 25.4&single&Y(5)&Y(14)&Y(8)&Y(9)&...(...)&Y\\
JH 223&4 40 49.51&+25 51 19.2&300&N(10)&Y(14)&Y(9)&Y(9)&N(4)&Y\\
IW Tau&4 41 04.71&+24 51 06.2&42&N(12)&N(14)&N(8)&...(...)&N(4)&N\\
ITG 33A&4 41 08.26&+25 56 07.5&...&Y(29)&Y(18)&Y(9)&...(...)&N(11)&Y\\
CoKu Tau/4&4 41 16.81&+28 40 00.1&7.8&N(19)&N(14)&Y(8)&...(...)&Y(4)&Y\\
IRAS 04385+2550&4 41 38.82&+25 56 26.8&single&Y(1)&Y(14)&Y(8)&Y(9)&Y(11)&Class I\\
2M04414565 A&4 41 45.65&+23 01 58.0&32&N(24)&N(2)&...(...)&...(...)&...(...)&...\\
LkHa 332 G1&4 42 05.49&+25 22 56.3&35&N(6)&N(14)&N(8)&...(...)&Y(4)&N\\
LkHa 332 G2&4 42 07.33&+25 23 03.2&34&N(10)&N(14)&N(8)&...(...)&N(4)&N\\
V955 Tau&4 42 07.77&+25 23 11.8&47&Y(17)&Y(7)&Y(8)&Y(9)&Y(4)&Y\\
DP Tau&4 42 37.70&+25 15 37.5&15.5&Y(1)&Y(14)&Y(8)&Y(9)&N(4)&Y\\
GO Tau&4 43 03.09&+25 20 18.8&single&Y(15)&Y(14)&Y(16)&Y(9)&Y(4)&Y\\
RX J0446.7+2459&4 46 42.60&+24 59 03.4&7.4&Y(30)&N(14)&N(9)&...(...)&N(11)&N\\
DQ Tau&4 46 53.05&+17 00 00.2&single&Y(12)&Y(14)&Y(8)&...(...)&Y(4)&Y\\
Haro 6-37&4 46 58.98&+17 02 38.2&48&Y(15)&Y(14)&Y(8)&...(...)&Y(4)&Y\\
DR Tau&4 47 06.21&+16 58 42.8&single&Y(12)&Y(14)&Y(8)&...(...)&Y(4)&Y\\
DS Tau&4 47 48.59&+29 25 11.2&single&Y(12)&Y(14)&Y(8)&...(...)&Y(4)&Y\\
UY Aur&4 51 47.38&+30 47 13.5&127&Y(17)&Y(14)&Y(8)&...(...)&Y(4)&Y\\
IRAS 04489+3042&4 52 06.68&+30 47 17.5&single&Y(1)&Y(14)&Y(3)&...(...)&...(...)&Class I\\
StHa 34&4 54 23.68&+17 09 53.5&171&Y(1)&...(...)&Y(31)&...(...)&N(4)&Y\\
GM Aur&4 55 10.98&+30 21 59.5&single&Y(12)&Y(14)&Y(25)&...(...)&Y(4)&Y\\
LkCa 19&4 55 36.96&+30 17 55.3&single&N(6)&N(14)&N(2)&...(...)&N(4)&N\\
SU Aur&4 55 59.38&+30 34 01.6&single&Y(32)&Y(14)&Y(8)&...(...)&Y(4)&Y\\
HBC 427&4 56 02.02&+30 21 03.8&4.7&N(10)&N(14)&N(8)&...(...)&N(4)&N\\
V836 Tau&5 03 06.60&+25 23 19.7&single&Y(1)&Y(14)&Y(8)&...(...)&Y(4)&Y\\
CIDA-8&5 04 41.40&+25 09 54.4&single&Y(33)&Y(14)&Y(2)&...(...)&Y(4)&Y\\
CIDA-9&5 05 22.86&+25 31 31.2&340&Y(33)&Y(14)&...(...)&...(...)&Y(4)&Y\\
CIDA-10&5 06 16.75&+24 46 10.2&12&N(33)&N(14)&...(...)&...(...)&N(4)&...\\
CIDA-11&5 06 23.33&+24 32 19.9&14.1&Y(33)&Y(14)&...(...)&...(...)&N(4)&Y\\
RX J0507.2+2437&5 07 12.07&+24 37 16.4&single&N(30)&N(14)&...(...)&...(...)&...(...)&...\\
RW Aur&5 07 49.54&+30 24 05.1&200&y(1)&Y(7)&Y(8)&...(...)&Y(4)&Y\\
CIDA-12&5 07 54.97&+25 00 15.6&single&Y(33)&Y(14)&...(...)&...(...)&N(4)&Y\\
 \enddata
 \tablenotetext{a}{Class I systems have near-infrared excesses, but since the emission can't be distinguished as coming from the envelope or a disk, we mark those systems here and omit them from our analysis.}
 \tablecomments{For each diagnostic, the number in parenthesese denotes the original source for the judgement. References: 
1) \citet{White:2004fb}; 
2) \citet{Luhman:2010cr};
3) \citet{Furlan:2008di}; 
4) \citet{Andrews:2005qf}; 
5) \citet{Luhman:2006dp};
6) \citet{White:2001jf}; 
7) \citet{McCabe:2006az}; 
8) \citet{Furlan:2006nl}; 
9) \citet{Rebull:2010xf}; 
10) \citet{Kenyon:1998ol};
11) \citet{Schaefer:2009xi};
12) \citet{Hartigan:1995tu};
13) \citet{Briceno:1998pz};
14) \citet{Hartmann:2005fd};
15) \citet{Hartmann:1998ly};
16) \citet{Najita:2007fr};
17) \citet{Hartigan:2003jy};
18) \citet{Luhman:2006tg};
19) \citet{Nguyen:2009vj};
20) \citet{Duchene:1999ux};
21) \citet{Hartmann:1991gf};
22) \citet{Espaillat:2007rq};
22) \citet{Muzerolle:2003xd};
23) \citet{Kraus:2009uq};
25) \citet{Calvet:2005xf};
26) \citet{Gomez:1992dt};
27) \citet{Luhman:2003fu};
28) \citet{White:2003xz};
29) \citet{Martin:2000dg};
30) \citet{Briceno:1999nm};
31) \citet{Hartmann:2005ve};
32) \citet{Edwards:2006ul};
33) \citet{Briceno:1993hc}; 
34) \citet{Wahhaj:2010fj}.
 }
 \end{deluxetable}

 \begin{deluxetable}{lcccl}
 \tabletypesize{\scriptsize}
 \tablewidth{0pt}
 \tablecaption{Circumbinary Disk Census of Other Nearby Associations}
 \tablehead{\colhead{Name} & \colhead{RA} & \colhead{DEC} & \colhead{Binary} & \colhead{Disk}
 \\
 \colhead{} & \multicolumn{2}{c}{(J2000)} & \colhead{Sep (AU)} & \colhead{Present?}
 }
 \startdata
\multicolumn{5}{l}{Ophiuchus ($\tau \sim 1 \pm 1$ Myr; Greene \& Meyer 1995; Wilking et al. 2005)}\\
RXJ1621.4-2332&16 21 28.8&-23 32 38.9&9.4(1)&N(19)\\
RXJ1624.8-2239&16 24 51.3&-22 39 32.5&6.0(1)&N(19)\\
Halpha 21&16 25 15.2&-25 11 54.1&23(2)&Y(19)\\
ROX 1&16 25 19.3&-24 26 52.1&34(3)&N(19)\\
RXJ1625.4-2346&16 25 28.6&-23 46 26.5&5.1(1)&N(19)\\
ROXs 5&16 25 55.8&-23 55 09.9&26(4)&N(19)\\
W05-4-28&16 26 01.6&-24 29 44.9&15(2)&N(19)\\
BKLT162643-241635&16 26 43.7&-24 16 33.3&16(2)&N(19)\\
VSSG 3&16 26 49.2&-24 20 02.9&35(2)&N(19)\\
VSSG 5&16 26 54.4&-24 26 20.7&21(2)&Y(19)\\
GY 156&16 26 55.0&-24 22 29.7&23(2)&N(19)\\
BKLT162658-244529 B&16 26 58.4&-24 45 31.8&12(5)&Y(19)\\
WL 4&16 27 18.5&-24 29 05.9&26(2)&Y(19)\\
VSSG 17&16 27 30.2&-24 27 43.5&36(6)&Y(19)\\
GY 410&16 27 57.8&-24 40 01.8&28(2)&N(19)\\
Halpha 59&16 28 09.2&-23 52 20.5&15(2)&N(19)\\
W05-1-35&16 28 32.6&-24 22 44.9&10(5)&Y(19)\\
ROXs 42 C&16 31 15.7&-24 34 02.2&23(3)&Y(19)\\
ROXs 43 B&16 31 20.2&-24 30 00.7&2(5)&Y(19)\\
\hline
\multicolumn{5}{l}{Cha-I ($\tau \sim 2 \pm 1$ Myr; Luhman 2004)}\\
T5&10 57 42.2&-76 59 35.7&29(7)&Y(20)\\
Hn 4&11 05 14.67&-77 11 29.1&39(7)&N(20)\\
T21&11 06 15.4&-77 21 56.8&26(7)&N(21)\\
CHXR28&11 07 55.9&-77 27 25.8&26(7)&N(21)\\
CHXR37&11 09 17.7&-76 27 57.8&15(7)&N(21)\\
CHXR40&11 09 40.1&-76 28 39.2&28(7)&N(21)\\
T46&11 10 07.0&-76 29 37.7&23(7)&Y(21)\\
2M1110-7722&11 10 34.8&-77 22 05.3&11(7)&N(21)\\
CHXR47&11 10 38.0&-77 32 39.9&32(7)&Y(21)\\
CHXR59&11 13 27.4&-76 34 16.6&27(7)&N(21)\\
CHXR62&11 14 15.7&-76 27 36.4&22(7)&N(21)\\
CHXR68A&11 18 20.2&-76 21 57.6&19(7)&...(...)\\
\hline
\multicolumn{5}{l}{Upper Sco ($\tau \sim 5 \pm 1$ Myr; Preibisch et al. 2002; Slesnick et al. 2008)}\\
GSC 06764-01305&15 35 57.8&-23 24 04.6&8(8)&N(22)\\
RXJ1549.3-2600&15 49 21.0&-26 00 06.3&24(9)&N(23)\\
RXJ1550.0-2312&15 50 05.0&-23 11 53.7&4(8)&N(22)\\
RXJ1550.9-2534&15 50 56.4&-25 34 19.0&19(8)&N(22)\\
ScoPMS013&15 56 29.4&-23 48 19.8&13(9)&N(22)\\
ScoPMS015&15 57 20.0&-23 38 50.0&18(8)&...(...)\\
ScoPMS017&15 57 34.3&-23 21 12.3&8(8)&Y(24)\\
RXJ1558.1-2405&15 58 08.2&-24 05 53.0&33(8)&N(22)\\
ScoPMS019&16 00 00.0&-22 20 36.8&4(8)&...(...)\\
RXJ1600.5-2027&16 00 31.4&-20 27 05.0&27(9)&N(22)\\
ScoPMS020&16 01 05.2&-22 27 31.2&28(9)&...(...)\\
RXJ1601.3-2652&16 01 18.4&-26 52 21.3&12(9)&N(22)\\
RXJ1601.7-2049&16 01 47.4&-20 49 45.8&30(9)&N(22)\\
RXJ1601.8-2445&16 01 51.5&-24 45 24.9&11(9)&N(23)\\
RXJ1601.9-2008&16 01 58.2&-20 08 12.2&6(8)&N(23)\\
RXJ1603.9-2031B&16 03 55.0&-20 31 38.4&18(9)&N(22)\\
RXJ1604.3-2130B &16 04 21.0&-21 30 41.5&12(9)&...(...)\\
ScoPMS027&16 04 47.8&-19 30 23.1&6(8)&N(22)\\
USco-160517.9-202420&16 05 17.9&-20 24 19.5&2(8)&N(22)\\
RXJ1607.0-2036&16 07 03.6&-20 36 26.5&27(8)&...(...)\\
USco-160707.7-192715&16 07 07.7&-19 27 16.1&15(8)&N(22)\\
GSC 06209-00735&16 08 14.7&-19 08 32.8&4(8)&N(22)\\
USco-161031.9-191305&16 10 32.0&-19 13 06.2&21(8)&N(22)\\
ScoPMS052&16 12 40.5&-18 59 28.3&21(10)&N(24)\\
GSC 06793-00819&16 14 11.1&-23 05 36.2&32(10)&Y(25)\\
\hline
\multicolumn{5}{l}{$\eta$ Cha ($\tau \sim 8 \pm 1$ Myr; Lyo et al. 2004; Lawson et al. 2009; Ortega et al. 2009)}\\
$\eta$ Cha 1&8 36 56.2&-78 56 45.5&13(11)&N(26)\\
$\eta$ Cha 9&8 44 16.4&-78 59 08.1&20(11)&Y(26)\\
$\eta$ Cha 12&8 47 56.8&-78 54 53.2&4(A)&N(26)\\
\hline
\multicolumn{5}{l}{TWA ($\tau \sim 8 \pm 2$ Myr; Song et al. 2003; Ortega et al. 2009)}\\
TWA-2&11 09 13.9&-30 01 40.0&21(12)&N(27)\\
HD 98800&11 22 05.3&-24 46 39.8&31(13)&Y(27)\\
TWA-5&11 31 55.4&-34 36 27.4&3(14)&N(27)\\
\hline
\multicolumn{5}{l}{BPMG ($\tau \sim 12 \pm 4$ Myr; Zuckerman et al. 2001; Ortega et al. 2009)}\\
GJ 3305&04 37 37.3&-02 29 28&3(15)&N(28)\\
HIP 23418&05 01 58.8&+09 59 00&31(16)&N(28)\\
CD-64$^o$ 1208&18 45 37.0&-64 51 45&5(18)&N(28)\\
 \enddata
 \tablecomments{For binary systems, the number in parenthesese denotes the original measurement of the binary separation. For the disk assessment, the number in parenthesese denotes the original source for the judgement. Since the data available are much more heterogeneous, we only base the judgement on 20--30$\mu$m data from IRS or MIPS. References: 
1) Kraus \& Ireland, in prep.;
2) \citet{Ratzka:2005nx};
3) \citet{Ghez:1993xh};
4) \citet{Ageorges:1997rr};
5) \citet{Simon:1995yi};
6) \citet{Costa:2000eu};
7) \citet{Lafreniere:2008fq};
8) \citet{Kraus:2008zr};
9) \citet{Kohler:2000lo};
10) \citet{Metchev:2009hh};
11) \citet{Kohler:2002bs};
A) \citet{Brandeker:2006cr};
12) \citet{Brandeker:2003wd};
13) HIPPARCOS \citep{Perryman:1997oq};
14) \citet{Konopacky:2007bh};
15) \citet{Kasper:2007od};
16) \citet{Delfosse:1999cy};
17) \citet{Prato:2002lk};
18) \citet{Biller:2007kj};
19) \citet{Padgett:2008kl};
20) \citet{Luhman:2008qf};
21) \citet{Luhman:2008hc};
22) \citet{Carpenter:2006hf};
23) \citet{Padgett:2006qe};
24) \citet{Carpenter:2009qe};
25) \citet{Dahm:2009cr};
26) \citet{Gautier:2008dq};
27) \citet{Low:2005lq};
28) \citet{Rebull:2008fu};
 }
 \end{deluxetable}


 \begin{deluxetable}{lccccr}
 \tabletypesize{\scriptsize}
 \tablewidth{0pt}
 \tablecaption{Disk Census of Other Nearby Populations}
 \tablehead{\colhead{Region} & \colhead{Age} & \colhead{Mass/SpT} & \colhead{$N_{disk}/N_{tot}$} & \colhead{$F_{disk}$} & \colhead{Refs}
 \\
 \colhead{} & \colhead{(Myr)} & \colhead{Range}
 }
 \startdata
NGC 1333&1&$\ga0.05M_{\sun}$&72/87&$83^{+3}_{-5}\%$&1\\
Taurus&2&G0-M4&83/120&$69^{+4}_{-5}\%$&2\\
Cha-I&2&G0-M3.5&56/85&$66^{+5}_{-6}\%$&3\\
NGC 2068/2071&2&G6-M6&53/67&$79^{+4}_{-6}\%$&4\\
IC 348&3&G0-M3.5&48/129&$37^{+4}_{-4}\%$&3\\
$\sigma$ Ori&3&0.25-2.0 $M_{\sun}$&64/153&$42^{+4}_{-4}\%$&5\\
Tr 37&4&G0-M2&81/166&$49^{-4}_{-4}\%$&6\\
$\gamma$ Vel&5&0.25-2.0 $M_{\sun}$&5/398&$1.3^{+0.8}_{-0.3}\%$&7\\
Upper Sco&5&G0-M4&17/126&$14^{+4}_{-2}\%$&8\\
TWA&8&F0-M4&4/15&$27^{+14}_{-8}\%$&9,10\\
$\eta$ Cha&8&F0-M4&3/10&$30^{+17}_{-10}\%$&11\\
NGC 7160&10&G0-M2&2/49&$4.1^{+5.0}_{-1.3}\%$&6\\
BPMG&12&F0-M4&0/22&$<$5\%&12\\
 \enddata
 \tablecomments{The samples were chosen to approximately match the 0.25--2.0 $M_{\sun}$ range of our Taurus sample. However, the available stars in each region sample slightly different ranges, and in cases where completeness corrections are required, we generally can not rebin the data to perfectly match our range. All disk identifications were made using observations at 10--30 $\mu$m. References: 
1) \citet{Gutermuth:2008tg}; 
2) Section 2; 
3) \citet{Luhman:2008qf}; 
4) \citet{Flaherty:2008dz};
5) \citet{Hernandez:2007zu}; 
6) \citet{Sicilia-Aguilar:2006fk}; 
7) \citet{Hernandez:2008ai};
8) \citet{Carpenter:2006hf}; 
9) \citet{Low:2005lq}; 
10) \citet{Plavchan:2009dp};
11) \citet{Gautier:2008dq};
12) \citet{Rebull:2008fu}.
 }
 \end{deluxetable}

\end{document}